\documentclass[iop,cha,twocolumn,showpacs]{revtex4}
\usepackage{amsfonts,amsmath,amssymb,latexsym,epsfig}

\usepackage{lipsum}

\usepackage{color,hyperref}
\hypersetup{
    colorlinks,%
    citecolor=blue,%
    filecolor=blue,%
    linkcolor=blue,%
    urlcolor=blue
}

\usepackage{color}
\definecolor{bluecolor}{rgb}{0,0.,1.}

\definecolor{redcolor}{rgb}{.7,0.,0.}

\newcommand{\pr}[1]{\left( #1\right)}
\newcommand{\prr}[1]{\left[ #1 \right]}
\newcommand{\es}[1]{\begin{equation}\begin{split}#1\end{split}\end{equation}}

\newcommand{\N}{\mathbb{N}}

\newcommand{\Z}{\mathbb{Z}}

\begin{document}

\title{Faster than expected escape for a class of fully chaotic maps}
\author{Orestis Georgiou$^{1}$, Carl P. Dettmann$^{2}$ and Eduardo G. Altmann$^{1}$}
\affiliation{1 Max-Planck-Institute for the Physics of Complex Systems, 01187, Dresden, Germany.\\
2 School of Mathematics, University of Bristol, BS8 1TW, Bristol, UK.}

\begin{abstract}
We investigate the dependence of the escape rate on the position of a hole placed in uniformly hyperbolic systems admitting a finite Markov partition.
We derive an exact periodic orbit formula for finite size Markov holes which differs from other periodic expansions in the literature and can account for additional distortion to maps with piecewise constant expansion rate.
Using asymptotic expansions in powers of hole size we show
that for systems conjugate to the binary shift, the average escape rate is always larger than the expectation based on the hole size.
Moreover, we show that in the small hole limit the difference between the two decays like a known constant times the square of the hole size.
Finally, we relate this problem to the random choice of hole positions and we discuss possible extensions of our results to non-Markov holes as well as applications to leaky dynamical networks.
\end{abstract}

\pacs{
05.45.Ac, 
05.60.Cd, 
02.50.Ga  
}

\maketitle

\textbf{
In deterministic chaotic systems, the probability that a particle does not escape through some pre-specified leaking region or hole decays exponentially with time.
One feature of this problem that has received recent attention is that even if the escape rate is of the order of the hole size it exhibits strong fluctuations depending on the hole's position.
In this paper we compare the typical escape rate obtained through two different procedures of introducing the holes in the setting of uniformly hyperbolic maps.
In the first case, a small hole is placed at random in a specific location of the phase space and so the expected escape rate is the average over all possible hole positions.
In the second case, the random choice of hole position is performed independently at each time step and trajectory.
This leads to a commonly used estimate for the average escape rate and corresponds also to the physical picture that the map is partially leaking in the whole phase space.
While both these averages are equal in the limit of small holes, we show analytically (and confirm numerically) that for small but finite sized Markov holes in systems conjugate to the binary shift, the former is greater than the latter and so escape is faster than expected.
}

\section{Introduction}

\begin{figure}[t!]
\begin{center}
\includegraphics[scale=0.21]{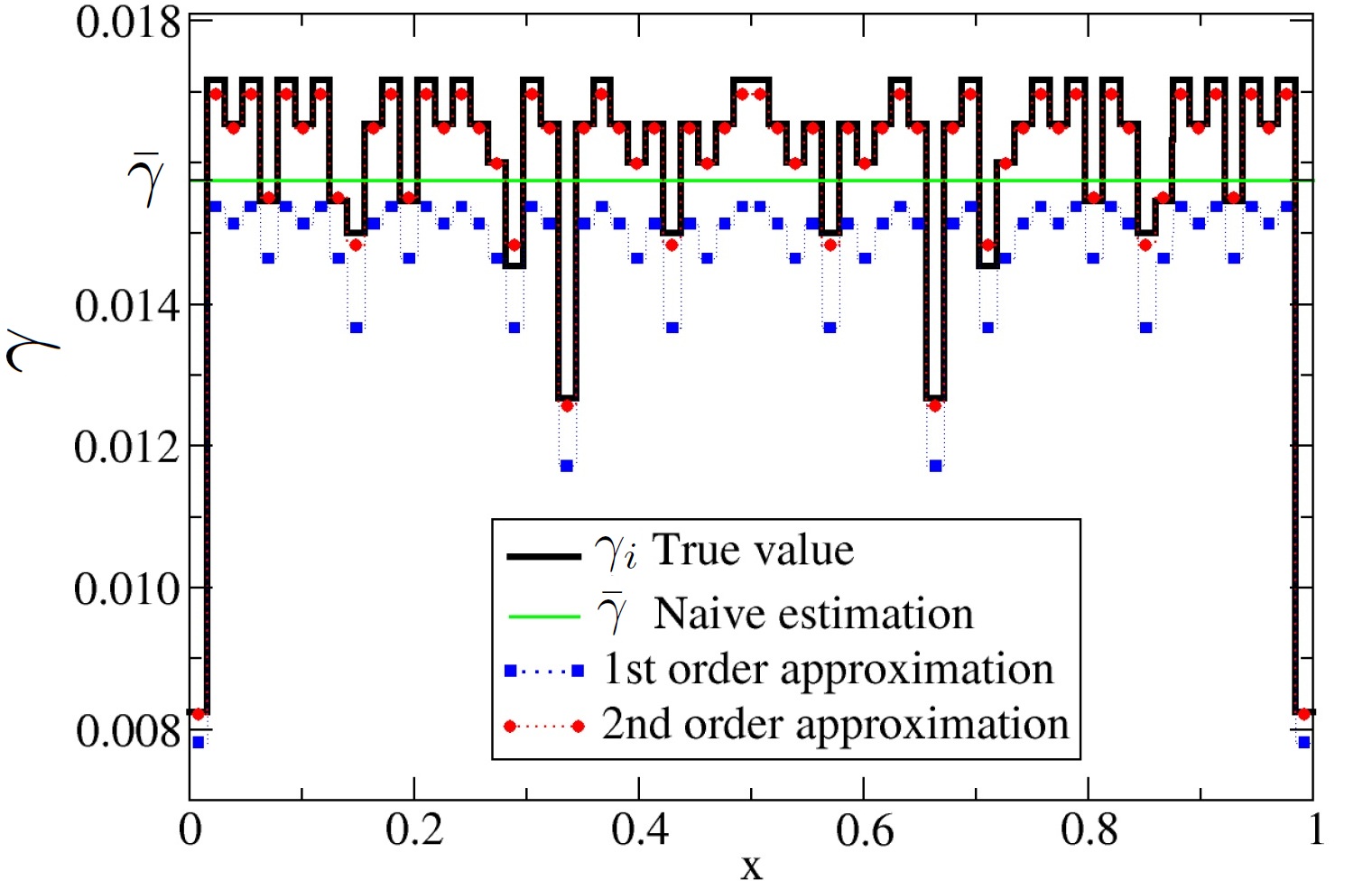}
\caption{\label{fig:n56} (Color online) The escape rate $\gamma$ of the open doubling map is plotted as a function of position for holes of size $h=2^{-n}$ and $n=6$ centered at $ih + h/2$ for $i=0,\ldots,2^{n}-1$.
The true values are indicated by the black full curve, the first order approximation of Eq.~\eqref{BY} by the blue squares, while the red dots correspond to the second order approximations calculated in Eqs.~\eqref{gammaA} and \eqref{gammaB}.
The width of each tower produced by the black full curve equals the size of the holes.
The horizontal green line indicates the naive estimate of Eq.~\eqref{naive}. }
\end{center}
\end{figure}

In recent years physical problems and mathematical results have motivated a renewed interest in the problem of placing holes through which trajectories can leak out from otherwise closed chaotic dynamical systems \cite{Yorke79,BB90} (for a recent review see Ref. \cite{APT12}).
The non-trivial aspect of this problem is that the properties of the open system depend sensitively on
the position of the hole \cite{PP97,AT09,KL09,AB10,BY11,DG11,KGDK12,DW12,FP12}.
For instance, in Fig.~\ref{fig:n56} we show the escape rate $\gamma_i$ (i.e. the exponential rate of decay of smooth initial conditions) of the fully chaotic one-dimensional
doubling map for different positions $i=0\ldots 2^{n}-1$ for holes of size $h=2^{-n}$ and $n=6$.
In the limit of small holes sizes $h \rightarrow 0$, it is well known that for any position (apart from a zero measure set)  $\gamma/h \rightarrow 1$.
The most natural (and naive) approximation to finite (but small) holes assumes that the (conditionally invariant) density \cite{DY06} remains uniform inside the open system and therefore
\es{
\bar{\gamma}=-\ln(1-h)= h + \frac{1}{2}h^{2}+ \mathcal{O}(h^{3})
\label{naive}.
}
The expectation $\bar{\gamma}$ appears as horizontal lines in Fig.~\ref{fig:n56} and correctly predicts the order of magnitude of the escape rate.
The most striking deviation from this general feature are the deep minima, which are located at the positions of the lowest order periodic orbits \cite{PP97}.
Considering a sequence of holes shrinking to a $\wp$-periodic point of the
open doubling map, Ref.~\cite{BY11} shows rigorously 
that the escape rate is to first order given by
\es{
\tilde{\gamma} = h(1-2^{-\wp})+ o(h). \label{BY}
}
Observing this estimation in Fig.~\ref{fig:n56} we see that it provides an improvement over the naive estimation of Eq.~\eqref{naive} with $\tilde{\gamma} < \bar{\gamma}$, and correctly captures some of the fluctuations in $\gamma_i$ for finite size holes.

In this paper we argue that there are many additional interesting features in the position dependence of~$\gamma_i$, apart from the minima, and require one to go beyond Eq. (\ref{BY}).
To see this, consider the average escape rate $\langle \gamma \rangle=2^{-n}\sum_{i=0}^{2^{n}-1} \gamma_i$ over the $2^n$ equally spaced
hole positions (which correspond also to the expected $\gamma_i$ if $i$ is chosen randomly).
From Eq.~\eqref{BY} we could expect that 
periodic orbits typically reduce the escape rate and therefore $\langle \gamma \rangle$  would be smaller than $\bar{\gamma}$.
In this paper we will show that for the doubling map and for a large class of fully chaotic Markov systems the opposite is true and
\es{
\langle \gamma \rangle \ge \bar{\gamma}
\label{inequality}
.}
We will show that inequality can be obtained through an asymptotic expansion to second order in hole size $h$ of an exact periodic orbit formula (see Eq. \eqref{poform} below). As seen in Fig.~\ref{fig:n56}, this expansion provides a much better estimation of the true values of ~$\gamma_i$ and can account for the many local maxima for which $\gamma_i /h >1$.

Our paper is structured as follows: In Sec.~\ref{sec.background} we introduce the formal definitions and background information needed for our
calculations. In Sec.~\ref{sec.results} we present the main results for the doubling map, including the new periodic orbit formula and the
demonstration of the inequality in Eq.~\eqref{inequality}. In Sec.~\ref{sec.extensions} we extend these results to other uniformly hyperbolic systems admitting a finite Markov partition including the two dimensional baker map.
In Sec.~\ref{sec.exception} we discuss the case of skewed maps and show how the inequality can be generalized or even broken.
Finally, in Sec.~\ref{sec.conclusions} we conclude and discuss the generality of our results, possible extensions to Non-Markov holes and applications to other research areas such as leaky dynamical networks.

\section{Formalities and background motivation}
\label{sec.background}

We  consider the function $\rho({\bf x}, n)$ describing the density of representative points in phase space $\mathcal{M}$ at time step $n$ which evolves under the action of a deterministic closed map $f:\mathcal{M}\rightarrow\mathcal{M}$.
The evolution of $\rho({\bf x}, n)$ can be understood best in the language of operators such that $\rho({\bf x},n)=\mathcal{L}^{n}\rho({\bf x},0)$,
where $\mathcal{L}$ is the Perron-Frobenius operator associated with $f$ \cite{BG97}. Thus, the normalized invariant density $\rho({\bf x})$ of $f$ is the eigenfunction related to the largest (in modulus) eigenvalue of $\mathcal{L}$.
In Ulam's method \cite{Ulam64}, $\mathcal{L}$ is approximated by an $N\times N$ \textit{transfer matrix} $T$ with elements corresponding to the transition probabilities between the $N$-partitioned phase space. There has been much work in recent years involving the convergence rates of Ulam's method \cite{GF07} as $N\to\infty$ as it can be used as a basis for rigorous computations \cite{WB11}.
For piecewise linear maps admitting a finite Markov partition such as the maps considered here,
the leading eigenfunction is piecewise constant, and so the treatment using a finite matrix is exact \cite{BG97}. 
Thus, such maps are called Markov maps.

In general, a closed map can be opened by choosing any subset of its phase space $\mathcal{M}$ as a hole through which trajectories can escape never to return.
For Markov maps, it is natural to consider the $i^{\textrm{th}}$ element of the Markov partition as the hole $H_i \subset\mathcal{M}$.
Here, we denote $\hat{f}$ as the open map corresponding to $f$.
At each iterate $j$ of $\hat{f}$, a proportion of mass $\nu_{j}$ may be lost to $H_i$ and therefore $\hat{f}: \mathcal{M}\setminus H_i \rightarrow \mathcal{M}$.
Typically, in strongly chaotic maps (e.g. with exponential decay of correlations), the proportion of mass remaining in the system decays exponentially with $n$ such that the escape rate
\es{\gamma = \lim_{n\rightarrow\infty} -\frac{1}{n}\ln P(n),
\label{eq6}}
is well defined. Here, $P(n)=\prod_{j=1}^{n}(1-\nu_{j})$ is called the survival probability.
The leading eigenfunction of the open Perron-Frobenius operator $\hat{\mathcal{L}}$ associated with $\hat{f}$
is now the \textit{conditionally} invariant density $\rho_{c}({\bf x})$ (see also Ref.~\cite{DY06}) such that
$\hat{\mathcal{L}}^{n}\rho_{c}({\bf x})=e^{- \gamma n}\rho_{c}({\bf x})$, and the escape rate can be obtained from the leading eigenvalue $\lambda<1$ of $\hat{\mathcal{L}}$ through the relation $\gamma= -\ln \lambda$ (see also Refs.~\cite{BFGTM12}).
For Markov maps, $\rho_{c}({\bf x})$ is again piecewise constant and so one looks at the leading eigenvalue of the finite matrix $T_{i}\equiv \hat{\mathcal{L}}$, where the subscript $i\in[0,N-1]$ characterizes the position of $H_i$ and is simply a zero-valued $(i+1)^{\textrm{th}}$ column.
Note that the matrices $T$ and $T_i$ are non-negative and hence by the Perron-Frobenius theorem
have a real positive eigenvalue (the Perron root) which is greater or equal in absolute value than all other eigenvalues.
Moreover, they are sparse and hence one can utilize preconditioned iterative solvers for fast computations.

For the greater part of this paper we will use the one dimensional doubling map $f(x)= 2x$ (mod $1$) as our paradigm example and later
generalize the results to other systems.
We choose the doubling map as it is uniformly expanding and has invariant density $\rho(x)=1$ on the unit interval, thus making it amenable to analysis with the methods we employ.
In particular, there is a close correspondence between the binary representation of a point $x=0.a_1 a_2 a_3 \dots_2$, and the symbolic dynamics for the partition $\{ [0,1/2) , [1/2,1] \}$, modulo minor details to do with dyadic rationals (i.e. fractions with denominator a power of $2$), a zero measure set.
Hence, with the partition $I_{n}=\{I_{n,i}\}_{i=0}^{2^{n}-1}$ where $I_{n,i}=[i,i+1]2^{-n}$ for $n>0$ and $i\in[0,2^{n}-1]$, one can consider the open doubling map $\hat{f}$ with a Markov hole $H_i =I_{n,i}$ of size $h=|H_i|=2^{-n}$ with respect to the relevant invariant measure of the closed map $f$
(in this case Lebesgue).
The leading eigenfunction of the corresponding $2^n\times2^n$ transfer matrix $T_i$ is piecewise constant on these intervals and so the leading eigenvalue can be calculated exactly as the root of a finite polynomial, or numerically to arbitrary precision.

It originally came as a surprise when the escape rate of uniformly hyperbolic systems was shown to be strongly dependent on the position of $H_i$, allowing for the possibility of escape through some holes to be as fast as through holes which are twice as big \cite{PP97}.
Similar results were also observed in two dimensional billiard models
\cite{BD07} (for a recent review see Ref.~\cite{D11}).
This striking observation (amongst others) was originally proved in Ref.~\cite{BY11} for finite size holes in a large class of hyperbolic maps, later generalized and applied to the context of metastability in Ref.~\cite{KL09}, and that of diffusion in Ref.~\cite{KGDK12}.
Significantly, for the open doubling map Eq.~\eqref{BY} suggests that the local escape rate for each hole is $\lim\limits_{h \rightarrow0} \gamma_i /h=1-2^{-\wp}$, and since the number of aperiodic points (i.e. irrational numbers) is of full measure, almost every point in $[0,1]$ has a local escape rate equal to $1$ \cite{BY11}.
Indeed, Eq.~\eqref{BY} which neglects the many local maxima where $\gamma_i /h >1$ (see Fig. \ref{fig:n56}) would wrongfully suggest that the $\gamma=h$ limit is approached from below since every finite sized hole contains a periodic orbit of finite period.
It was however noted in \cite{BY11} that $\gamma_i /h >1$ can occur when the shortest periodic orbit contained in a finite sized hole $H_i$ is maximal among all other possible holes of equal size.
Based on this observation, we will show here that for many holes and indeed for the average $\langle \gamma \rangle$, the $\gamma=h$ limit is in fact approached from above.

A commonly used and very useful estimate for $\gamma$ is given by $\bar{\gamma}$ as in Eq.~\eqref{naive}, which is what Ref.~\cite{AT09} refers to as the ``naive'' estimate since it is equivalent to a binomial estimate for recurrence times in a purely random process and also to the escape rate in the presence of strong noise \cite{AE11}.
Interestingly, a connection can also be made with random maps. We briefly recall that a random map is a discrete time process in which one of $N$ maps is selected at random and applied separately at each time step \cite{SP84}.
Thus, in the spirit of Ref.~\cite{LPPV96}, one can define an average transfer matrix for open Markov maps as
$\bar{T}= \sum_{i=0}^{N-1}q_{i} T_{i}$, with probability weights $q_{i}$.
The corresponding average escape rate is calculated from the leading eigenvalue of $\bar{T}$.
For the open doubling map (where $N=2^{n}$), if all but one $q_{i}$ are equal to zero, then we are in the situation described previously with the escape rate being strongly position dependent.
If however all transfer matrices are equally probable to occur then all $q_{i}=2^{-n}$, and $\bar{T}= (1-2^{-n})T$ where $T$ is the transfer matrix for the closed system. This case corresponds to {\em partial} leakage (reflection coefficient equals to $1-h$) uniform in the phase space and so the escape rate~$\bar{\gamma}$ equals that of Eq.~\eqref{naive} with $h=2^{-n}$
since $T$ is measure preserving and so has largest eigenvalue equal to $1$.

We are interested in the following natural question: \textit{If a Markov hole is chosen at random, what escape rate should one expect?}
or equivalently, \textit{What is the average of the curve shown in Fig. \ref{fig:n56}?}
We thus define this average for a Markov map with $N$ possible hole positions as
\es{
\langle \gamma \rangle= \sum_{i=0}^{N-1}h_{i} \gamma_{i}\label{expected}.
}
For the doubling map, all $h_{i}=2^{-n}$ and therefore $\langle \gamma \rangle$ is just an arithmetic mean. In the limit of large $n$ we have that $\langle \gamma \rangle$ converges to the size of the hole in agreement with both Eqs.~\eqref{naive} and \eqref{BY}.
It is the intention of this paper however to show that for a large class of Markov maps this limit is actually approached from above rather than below.
We do this by first deriving an exact periodic orbit formula for Markov maps (Sec.~\ref{ssec.pof}) which we then asymptotically expand to second order in hole size (Sec.~\ref{ssec.sha}). We then take the arithmetic mean of all $2^n$ escape rates and obtain an asymptotic expansion of $\langle \gamma \rangle$ (Sec.~\ref{ssec.exp}) leading to the inequality \eqref{inequality} for $n>1$.
Thus the escape rate is typically faster than the expected naive estimate $\bar{\gamma}$.

\section{Main results}
\label{sec.results}

\subsection{Periodic orbit formula for Markov maps}\label{ssec.pof}

As discussed above, the leading eigenvalue $\lambda$ for the open doubling map is given by the solution of a
polynomial equation, the characteristic equation of the transfer matrix $T_{i}$. To make progress we
need an explicit expression for this polynomial.  While this has been discussed in a number
of contexts, the form in which we express this, a new periodic orbit formula, is of interest
in its own right.

Historically, this problem has been considered from the point of view of the waiting time
distribution for finding a given fixed sequence $S$ of $n$ symbols in a sequence $\mathfrak{S}$ built drawing an independent identically distributed random variable,
which has practical applications in computer search algorithms and DNA sequence analysis.
While similar expressions were stated as early as 1966~\cite{Solovev66}, the most convenient
starting point is Thm 2.1 of Ref.~\cite{BT82} which gives (using our notation) the
following recursion relation for the probability $w_{t}$ of stopping after exactly $t\geq n$ symbols:
\es{
w_t=2^{-n}-2^{-n}\sum_{p=n}^{t-n}w_p -\sum_{p=t-n+1}^{t-1}w_p 2^{-(t-p)}\chi_{p+n-t}
\label{stop}
,}
where $\chi_r$ is one if the first and last $r$ symbols of $S$ are identical, otherwise
zero.

Since $\mathfrak{S}$ is constructed from independent random events (e.g. a toss of coin),
we expect $w_t$ to decay exponentially for large $t$ and therefore we assume that $w_t=c\lambda^t(1+o(t^{-\alpha}))$ for any power $\alpha>0$.
This assumption also reflects the exponentially decaying survival probability of strongly chaotic maps (cf. Eq. \eqref{eq6}) where the mixing property of the dynamics causes the system to `forget' its initial state and therefore $P(n)$ decays as a Poisson process.
Moreover, since the $w_t$ are probabilities they sum to one.
Thus, by the use of the formula for a geometric series, the first two terms on the RHS become
\es{
2^{-n} \left(1-\sum_{p=n}^{t-n}w_p \right)&= 2^{-n}\sum_{p=t-n+1}^\infty w_p \\ 
&=\frac{(2\lambda)^{-n}}{1-\lambda}c \lambda^{t+1}(1+o(t^{-\alpha}))
.}
Substituting this into Eq.~\eqref{stop}, writing $J=t-p$ in last sum on the RHS of Eq.~\eqref{stop} and relabeling $J$ as $p$, dividing everything by $c \lambda^{t}$ and then taking the limit $t\rightarrow\infty$ we arrive at an exact formula
\es{
1=\frac{\lambda}{1-\lambda}(2\lambda)^{-n} - \sum_{p=1}^{n-1}(2\lambda)^{-p}\chi_{n-p}
\label{eq.total}.
}

We now connect this problem to the characteristic equation of the transfer matrix $T_{i}$ (not restricted to the $T_i$ of the doubling
map). First we notice that for each hole indexed by $i=0 \ldots 2^{n}-1$, there is a fixed point of the map $f^n$.
We then interpret the symbolic sequence $S_i$ of the periodic orbit associated to this fixed point as the symbolic sequence of the hole $H_i$.
Significantly, in the case of the doubling map, $S_i$ is precisely the $n$-digit binary representation of the hole index $i$.
Next we notice that $\chi_{n-p}$ used above indicates exactly the number (zero or one) of periodic points of length $p$ in the interval $H_i=I_{n,i}$.
Thus Eq.~(\ref{eq.total}) can be written as
\es{
1=\frac{\lambda}{1-\lambda}(2\lambda)^{-n}
- \sum_{p=1}^{n-1} \sum_{\substack{{\bf x}:f^{p}({\bf x})={\bf x} \\ {\bf x}\in H_i}}(2 \lambda)^{-p}
.\label{eq.total++}
}

For the doubling map with Markov holes $H_i$, all of size $h=2^{-n}$, Eq.~\eqref{eq.total++} is exactly equal to the characteristic equation of $T_i$ multiplied by $h\lambda/(\lambda -1)$, and can be rearranged and expressed as
\es{
(\lambda-1)\pr{ 2^{n}\lambda^{n-1} + \sum_{p\in\mathcal{P}} (2^{n-p} \lambda^{n-p-1})} +1=0
\label{poconv}
,}
where $\mathcal{P}\subset[1,n-1]$ is the set of periods $p<n$ for periodic points in $H_i$.
We shall use this expression in the next subsection in order to obtain improved asymptotics in hole size formulas for the escape rate.

A more compact and perhaps elegant version of expression \eqref{eq.total++} can be obtained by noting that each period $p\geq n$ has exactly $2^{p-n}$ periodic points in the interval $H_i$ irrespective of the value of $i$.  Thus we have a geometric series
\es{
\sum_{p=n}^{\infty} \sum_{\substack{{\bf x}:f^{p}({\bf x})={\bf x} \\ {\bf x}\in H_i}}(z/2)^{p}= \sum_{p=n}^{\infty}2^{p-n}(z/2)^{p}=\frac{(z/2)^{n}}{1-z},
\label{Z10}}
which converges for $|z|<1$ and has an analytic continuation at all complex $z\neq 1$. We combine Eq. \eqref{Z10} with the finite sum of short periodic orbits in Eq. \eqref{eq.total++} to construct a function
\es{
Z_i(z)=1+ \sum_{p=1}^\infty \sum_{\substack{{\bf x}:f^{p}({\bf x})={\bf x} \\ {\bf x}\in H_i}} (z/2)^{p} \label{poform}
}
which can likewise be uniquely defined for all $z\neq 1$ by analytic continuation.
Thus we come to our first main result:
\emph{For the doubling map with a Markov hole $H_i$, we have that $Z_i(\lambda^{-1})=0$.}

Note that Eq. \eqref{poform} differs from other periodic expansions in the literature~\cite{AAC90,chaos} in that it enumerates periodic points in the hole rather than periodic orbits that avoid it (see also~\cite{AT09}), the sum is over all periodic points at all periods, and it is always divergent (i.e. requires analytic continuation for its definition). As with similar number-theoretic and dynamical zeta functions (see also Ref. \cite{chaos}), it has the paradoxical property that the (correctly interpreted) sum of a positive divergent series is zero. The derivation naturally generalizes to Markov maps with more than two full branches, including where each branch has a different expansion factor (i.e. skewed maps); here the factor $2^{-p}$ is replaced by the expansion factor of the relevant periodic orbit.  We emphasize that this periodic orbit formula is exact, and applies to Markov holes of all sizes.

\subsection{Small hole asymptotics}
\label{ssec.sha}

We now expand the escape rate in powers of the hole size $h$; as can be seen from a similar calculation~\cite{D11b} this can be useful whether or not the function is smooth. As in Ref.~\cite{D11b} we need to allow the coefficients to contain polynomial functions of $n=|\log_2 h|$.
We write $\gamma= \gamma^{(1)} h + \gamma^{(2)} h^{2} +\ldots $, so that the small hole expansion for the corresponding eigenvalue is of the form $\lambda= 1 - \lambda^{(1)} h - \lambda^{(2)} h^{2} $ up to second order such that $\gamma^{(1)}=\lambda^{(1)}$ and $\gamma^{(2)}= ( \lambda^{(2)}+  (\lambda^{(1)})^{2}/2 )$.
Substituting these into Eq.~\eqref{poconv} we obtain
\begin{widetext}
\es{
\sum_{p\in\mathcal{P}_0}\Bigg[ \bigg(1-(n-p-1)\lambda^{(1)}h &-\pr{(n-p-1)\lambda^{(2)}-\frac{(n-p-1)(n-p-2)}{2}(\lambda^{(1)})^{2} }h^{2}-\ldots \bigg) \\
&\times2^{n-p}\pr{-\lambda^{(1)} h -\lambda^{(2)} h^{2} -\ldots}\Bigg]+1=0
,}
\end{widetext}
where $\mathcal{P}_0= \mathcal{P}\cup\{ 0\}$.
Collecting terms of order $h^{0}=1$ and $h^{1}=2^{-n}$ gives
\es{\gamma^{(1)} = \pr{ \sum_{p\in\mathcal{P}_0}2^{-p} }^{-1}
\label{gamma1}
,}
\es{\gamma^{(2)} = \frac{ \sum_{p\in\mathcal{P}_0} (n-p-1/2)2^{-p}  }{ \pr{\sum_{p\in\mathcal{P}_0}2^{-p}}^{3} }
\label{gamma2}
,}
respectively. The expansion can be performed nicely to arbitrary order.

We now make the important observation, that each Markov hole may be one of two types.
Type A holes contain a single primitive periodic orbit of period $\wp\in[1, n/2]$ which is repeated up to $n-1$, and may also contain periodic orbits with periods $p\in(n/2,n-1]$ which are non-repeats of $\wp$.
Typically, type A holes are associated with escape rates $\gamma_i \lesssim h$ due to the short periodic orbits.
Type B holes contain none or many primitive periodic orbits of periods $p\in(n/2,n-1]$ and usually have escape rates $\gamma_i \gtrsim h$.
In particular, the maximal escape rates appear in type B holes and the number of such holes is given by the integer sequence~\cite{OEIS} $A003000$ which refers to the number of ``bifix-free'' words of length $n$ over a two-letter alphabet and increases like $\sim 0.2678 \times 2^{n}$.
We call the sets of type A and B holes $\mathcal{A}, \mathcal{B}\subset[1,2^{n}]$ respectively such that $|\mathcal{A}|+|\mathcal{B}|=2^{n}$.
This simple classification of holes turns out to efficiently capture the fluctuations of $\gamma_i$ and is therefore key in calculating the average given by Eq. \eqref{expected}.
We will now treat each case separately.

\subsubsection{Type A holes}

As there is only a single primitive periodic orbit of period $\wp\leq n/2$, the sums in \eqref{gamma1} and \eqref{gamma2} must count repeats of this orbit up to $n-1$,
that is
\es{
\sum_{p\in\mathcal{P}}2^{-p} &= \sum_{i=1}^{m} 2^{-i\wp} + \sum_{\substack{ p: \wp \nmid p \\ p \in \mathcal{P} }} 2^{-p} \\
& = 2^{-\wp}\frac{1-2^{-m\wp}}{1-2^{-\wp}} + \sum_{\substack{ p: \wp \nmid p \\ p \in \mathcal{P} }} 2^{-p}
,}
where $\wp=\textrm{min}(\mathcal{P})\leq n/2$ and $m= \lfloor \frac{n-1}{\wp} \rfloor \geq1$.
Substituting back into \eqref{gamma1} we obtain that
\es{\gamma^{(1)} = 1-2^{-\wp}+ 2^{-(m+1)\wp}-  \sum_{\substack{ p: \wp \nmid p \\ p \in \mathcal{P} }} 2^{-p}  \ldots
\label{gam1}}
Notice that the leading order term is equal to Eq.~\eqref{BY} since the sum in \eqref{gam1} is at most of order $\sim 2^{-n/2}$.

We now consider the numerator of $\gamma^{(2)}$ in \eqref{gamma2}. We have that
\es{
&\sum_{p\in\mathcal{P}_0} \pr{n-p-\frac{1}{2} }2^{-p}  \\
&=\sum_{i=0}^m \pr{n-\wp i- \frac{1}{2} }2^{-\wp i} + \sum_{\substack{ p: \wp \nmid p \\ p \in \mathcal{P} }} \pr{n-p-\frac{1}{2} }  2^{-p} \\
&= \frac{n-1/2}{1-2^{-\wp}} - \frac{\wp 2^{-\wp}}{(1-2^{-\wp})^{2}} + \sum_{\substack{ p: \wp \nmid p \\ p \in \mathcal{P} }} \pr{n-p-\frac{1}{2} }  2^{-p} \ldots
\label{num}
}
Approximating the denominator of $\gamma^{(2)}$ by $(1-2^{-\wp})^{-3}$, ignoring the sum (which is of higher order) and combining with \eqref{num}, we obtain that
\es{
\gamma^{(2)} = \pr{n-\frac{1}{2}} (1-2^{-\wp})^{2} - \wp 2^{-\wp}(1-2^{-\wp}) +\ldots
}

\subsubsection{Type B holes}

As there may be many primitive periodic orbits of length $p>n/2$ we expand Eq.~\eqref{gamma1} in a binomial series to get
\es{\gamma^{(1)} &= \pr{ 1+ \sum_{p\in\mathcal{P}}2^{-p} }^{-1}\\
&=  1- \sum_{p\in\mathcal{P}}2^{-p} + \pr{ \sum_{p\in\mathcal{P}}2^{-p}}^{2} -\ldots
.}
For $\gamma^{(2)}$ it is sufficient to keep only the leading order term given by $p=0$ such that $\gamma^{(2)}= n-1/2$.

\subsubsection{Comparison with true escape rate}

We define $\gamma_i$ as the escape rate corresponding to hole $H_i=I_{n,i}$ such that we have the following approximations for type A holes
\es{ \gamma&_{i\in\mathcal{A}}= \Big(1-2^{-\wp}+ 2^{-(m+1)\wp}-  \sum_{\substack{ p: \wp \nmid p \\ p \in \mathcal{P} }} 2^{-p} \Big) h \\
&+ \Big[(n-\frac{1}{2}) (1-2^{-\wp})^{2} - \wp 2^{-\wp}(1-2^{-\wp})\Big] h^{2}+ o(h^{2})
\label{gammaA}
,}
and for type B holes
\es{ \gamma_{i\in\mathcal{B}}= \prr{1- \sum_{p\in\mathcal{P}}2^{-p} }h + \prr{n-\frac{1}{2}} h^{2} +o(h^{2}).
\label{gammaB}}

Fig. \ref{fig:n56} shows a comparison of Eq.~\eqref{BY} and Eqs.~\eqref{gammaA} and \eqref{gammaB} against the true values of $\gamma$ for $n=6,7$. It is clear that the improved asymptotics now successfully capture both local maxima and minima of the escape rate.

\subsection{The average escape rate}\label{ssec.exp}

Equipped with Eqs.~\eqref{gammaA} and \eqref{gammaB} we can now take the arithmetic mean of all $\gamma_i$'s to obtain an asymptotic approximation to $\langle \gamma \rangle$.
To do this we need to know the number of periodic points $a(\wp)$ in $[0,1]$ with primitive orbit lengths $\wp\geq1$.
For the doubling map $f$ , we have that $a(\wp)= 2, 2, 6, 12, 30, 54,\ldots$ which exactly equals the number of aperiodic binary strings of length $\wp$ and is given by the integer sequence $A027375$  described by the formula\cite{OEIS}
\es{a(\wp)= \sum_{d|\wp}\mu(d)2^{\wp/d}
,}
where $\mu(d)$ here is the number theoretic M\"{o}bius function. Hence, there are $|\mathcal{A}|= \sum_{\wp=1}^{n/2}a(\wp)$ type A holes and $|\mathcal{B}|=2^{n}-|\mathcal{A}|$ type B holes.

We first sum over the escape rates of type A holes. For this it is sufficient to keep just the leading order term of $\gamma_{i\in\mathcal{A}}=(1-2^{-\wp})h$.
We have that
\es{
\sum_{i\in\mathcal{A}}\gamma_{i} &=  \Bigg( |\mathcal{A}| - \sum_{\wp=1}^{n/2}a(\wp)2^{-\wp} \Bigg) h \\
&=  \Bigg( |\mathcal{A}|- \sum_{d=1}^{n/2} \sum_{j=1}^{n/(2d)} \mu(d)2^{j(1-d)}\Bigg) h\\
&=  \Bigg( |\mathcal{A}| - \left\lfloor\frac{n}{2}\right\rfloor +  \sum_{d=2}^{n/2} \mu(d) \frac{2^{-k(d-1)}-1  }{1-2^{d-1}} \Bigg) h \\
&=  \Bigg(|\mathcal{A}| - \left\lfloor\frac{n}{2}\right\rfloor + \Big[ 2^{-\lfloor n/4\rfloor}+ \frac{2^{-2\lfloor n/6\rfloor}}{3}+\ldots\Big] \\
&+\sum_{d=2}^{n/2} \frac{-\mu(d)}{1-2^{d-1}}\Bigg) h
,\label{avA}
}
where we have set $\wp=j d$ and $k=\lfloor n/(2d) \rfloor$ in order to exchange the order of the sums in the second equality and
  $\lfloor x \rfloor$ is the integer part of $x$ (floor function).
Therefore, \eqref{avA} converges exponentially to
\es{\sum_{i\in\mathcal{A}}\gamma_i=  \pr{|\mathcal{A}| - \left\lfloor\frac{n}{2}\right\rfloor + \kappa} h
,\label{sgammaA}}
where $\kappa=\sum_{d=2}^{n/2} \frac{-\mu(d)}{1-2^{d-1}}\approx 1.382714$ for $n\gg1$.

We now sum over the escape rates of type B holes. Here it is necessary to keep terms up to second order in $h$.
We have that
\es{
&\sum_{i\in\mathcal{B}}\gamma_{i} =\sum_{i\in\mathcal{B}}\prr{ \Bigg( 1- \sum_{p\in\mathcal{P}} 2^{-p} \Bigg) h + \pr{n-\frac{1}{2}}h^{2}}\\
&=  \bigg( |\mathcal{B}| - \sum_{p=\lfloor n/2\rfloor+1}^{n-1} a(p)2^{-p} \bigg)h +  \pr{n-\frac{1}{2}}|\mathcal{B}| h^{2}\\
&=  \pr{|\mathcal{B}| - \pr{\left\lceil \frac{n}{2}\right\rceil-1} -\prr{2^{-\lfloor n/4\rfloor}-\ldots } }h \\
&+ \pr{n-\frac{1}{2}}|\mathcal{B}|h^{2}.
\label{avB}}
Therefore, \eqref{avB} converges exponentially to
\es{\sum_{i\in\mathcal{B}}\gamma_i=  \pr{|\mathcal{B}| + \left\lfloor \frac{n}{2}\right\rfloor + \frac{1}{2} } h
,\label{sgammaB}}
since $|\mathcal{B}|\sim 2^{n}-2^{\frac{n}{2}+1}$ for $n\gg1$.

Using \eqref{sgammaA} and \eqref{sgammaB} we may now calculate the average (mean) escape rate
\es{\langle\gamma \rangle &=2^{-n}\sum_{i=0}^{2^{n}-1}\gamma_i = 2^{-n}\pr{ \sum_{i\in\mathcal{A}}\gamma_i+\sum_{i\in\mathcal{B}}\gamma_i}\\
&= 2^{-n} +\pr{\kappa +\frac{1}{2}} 2^{-2n} +\ldots
.
\label{kappa}}
Note that the terms proportional to $n$ have exactly cancelled between the $\cal A$ and $\cal B$
contributions, thus the logarithmic terms suggested by the results of~\cite{D11b} do not appear,
at least to this order.

\begin{figure}[t]
\begin{center}
\includegraphics[scale=0.35]{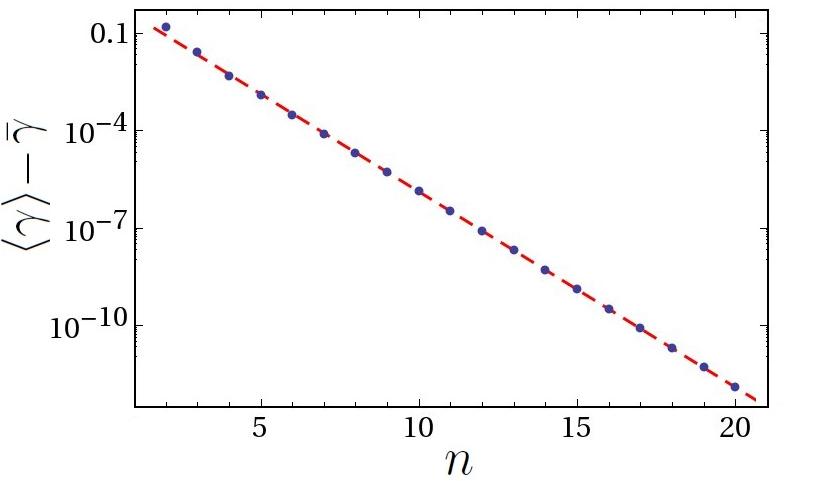}
\caption{\label{fig:compare} (Color online) The difference between $\langle\gamma\rangle$ and $\bar{\gamma}$ was computed numerically using long double (19 digit) precision for $n\leq20$ (blue dots). The red dashed line is given by the prediction $\langle\gamma\rangle-\bar{\gamma}=\kappa 2^{-2n}$. }
\end{center}
\end{figure}

Since $\kappa>0$, the above calculations show that for small but finite Markov holes, $\bar{\gamma} \leq \langle\gamma\rangle$ with equality only in the limit of $n\rightarrow\infty$. The difference between the two is $\langle\gamma\rangle-\bar{\gamma}=\kappa 2^{-2n}$ and is shown in Fig. \ref{fig:compare} on a log-linear scale.
Even if the derivation of Eq. \eqref{kappa} does not provide a rigorous proof of inequality \eqref{inequality} -- we have not properly bound Eqs. \eqref{gammaA} and \eqref{gammaB} -- the excellent numerical agreement for $n>3$ observed in Fig. \ref{fig:compare} strongly suggests that fluctuations due to higher order terms are negligible and that \eqref{inequality} is true for Markov holes of all sizes.

We now briefly summarize the results presented in this section.
We first obtained a periodic orbit formula \eqref{poform} for the leading eigenvalue $\lambda$ of the open transfer matrix $T_i$ of uniformly hyperbolic maps admitting a finite Markov partition.
Using the doubling map as our main example we have divided all holes into two distinct groups which correspond to escape rates typically greater or smaller than the expected naive estimate \eqref{naive}, and also obtained improved asymptotic formulas for them (see Eqs. \eqref{gammaA} and \eqref{gammaB}) to second order in hole size.
Taking the arithmetic mean of all different escape rates, we attained an asymptotic expansion for $\langle \gamma \rangle$, thus analytically showing that escape in the binary shift is faster than expected (see Eq. \eqref{inequality}) for finite size holes.
We now consider generalizations of these results to other open dynamical systems.

\section{Generalizations and extensions to other maps}\label{sec.extensions}

\begin{figure*}
\includegraphics[scale=0.34]{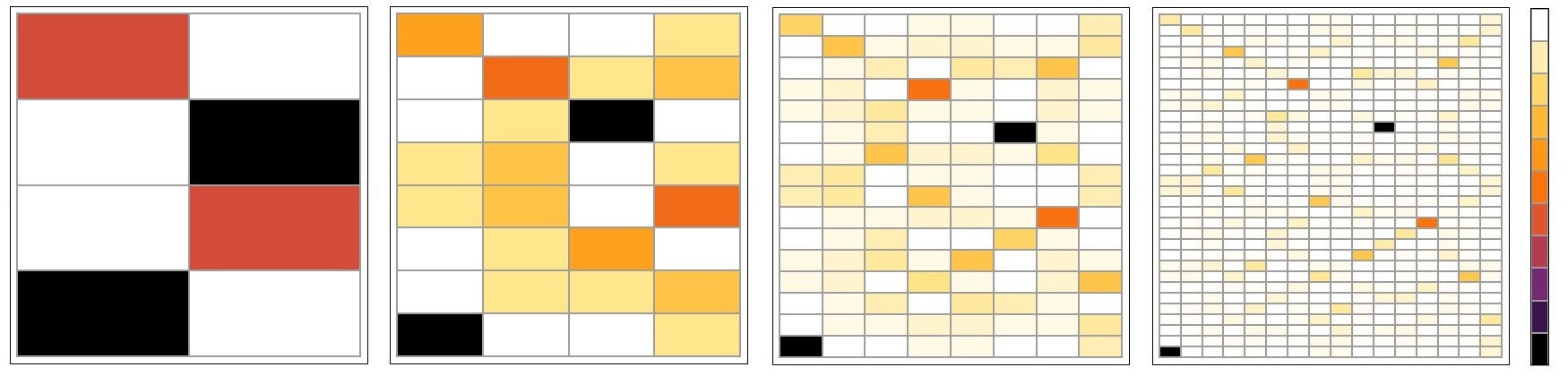}
\caption{\label{fig:baker} (Color online) Plot of $\gamma_{k}$ for the baker map \eqref{bake} for Markov holes $h= |I_{m,n,i,j}|$ for $m=n-1$ and $n=2,3,4$ and $5$. The colors have been scaled in each case so that the minimum and maximum escape rates are represented by black and white boxes respectively (see color scale on the right). }
\end{figure*}

\subsection{Linear expanding maps}
The above results can be easily generalized for linear expanding maps on the interval with uniform invariant densities of the form $x\mapsto
s x$ (mod $1$), $s\in \Z$ and $|s|>1$. This can be done by using $|s|$ instead of $2$ in Eq.~\eqref{poconv}, and expanding $\lambda$ in terms of the size of the new Markov holes $h=|s|^{-n}$.
We remark that maps with $s<0$ have been recently used in the context of polygonal billiards with (non-conservative) ``pinball'' type dynamics to rigorously prove hyperbolicity \cite{MPS10}. More specifically, the case of $s=-2$ corresponds to the so called ``slap'' map of an equilateral triangle billiard.

\subsection{Tent map}

The tent map
\es{f(x)=1-2\left|x-\frac{1}{2}\right|, \quad \text{for  $x\in[0,1]$}
\label{tent}}
stretches $[0,1]$ to twice its original length and then folds it in half back onto $[0,1]$.
Although the map has a uniform invariant density, the dynamics no longer commutes with the symmetry $x\rightarrow 1-x$.
Nevertheless, our results also apply here since the tent map is a metric conjugacy onto the left shift symbolic space and so shares the same binary symbolic dynamics and hierarchy of periodic sequences as the doubling map.

\subsection{Baker map}

The two dimensional baker map
\es{f(x,y)= \begin{cases}
(2x,y/2), &   \text{for} \quad 0\leq x<1/2 ,\\
(2-2x,1-y/2), &   \text{for} \quad 1/2\leq x<1 , \end{cases}
\label{bake}}
is area preserving with respect to Lebesgue (i.e. $\rho(x,y)=1$), and is a two-dimensional analog of the tent map.
The unit square is squeezed uniformly two times in the vertical $y$ direction and stretched in the horizontal $x$ direction. It is then cut in half, and the right half is folded over and placed on top of the left half.
It is thus topologically conjugate to the Smale horseshoe map.
Note that if the right half is not folded over, but simply placed (un-rotated) on top the map would be analogous to the doubling map.
Unlike all previously mentioned maps however the baker map is invertible and so can mimic chaotic dynamics in 
Hamiltonian systems.
The quantum version of the map has been used to explore the classical to quantum correspondence in the semiclassical limit \cite{BV87}, and to study the emergence of fractal Weyl laws in the quantum theory of open systems \cite{NM05}.

As with linearly expanding maps, the binary representations of points $x$ and $y$ in $[0,1]^2$ have a close correspondence with the symbolic dynamics for the partition $\{ [0,1/2) , [1/2,1]\} \times \{[0,1/2) , [1/2,1]\}$.
Therefore, unlike  toral automorphisms (including for example Arnold's cat map), the Baker map shares the same hierarchy of periodic sequences as the doubling map.
Moreover, the eigenfunctions of $\mathcal{L}^n$ are piecewise constant on the rectangles given by
\es{
I_{m,n,i,j}=\prr{\frac{i}{2^{m}}, \frac{i+1}{2^{m}}}\times \prr{\frac{j}{2^{n}}, \frac{j+1}{2^{n}}}
,}
for $i\in[0,2^{m}-1]$, $j\in[0,2^{n}-1]$, and $n,m\in\N^{+}$.
The map is Markov and thus the evolution of densities can be exactly described by a $2^{n+m}\times 2^{n+m}$ transfer matrix $T$.

The baker map is opened by choosing a Markov hole $H_{i,j}=I_{m,n,i,j}$ of size $h=2^{-(n+m)}$ and setting the $k^{th}$ column of $T$ equal to zero and denoting the open transfer matrix as $T_k$.
Fig. \ref{fig:baker} shows the variation of $\gamma_{k}$ as a function of position for $m=n-1$ and $n=2,3,4,$ and $5$.
Since the leading eigenvalue of $T_{k}$ is a solution to Eq.~\eqref{poform}, the small hole asymptotic formulas \eqref{gammaA} and \eqref{gammaB} of section 4.3 hold. Moreover, for $n\gg1$ we have that $\langle \gamma \rangle - \bar{\gamma} = \kappa 2^{-2(m+n)}$. We have confirmed this numerically (not shown here) for $n\leq 10$.

\subsection{Logistic map}

There also exist examples with nonlinear dynamical equations and non-uniform invariant densities for which the above results hold.
One such example is the logistic map
\es{f(x)=4x(1-x), \quad \text{for  $x\in[0,1]$}
\label{logistic}}
with a partition of the interval given by
\es{I_{n,i}= \prr{\sin^2 \pr{\frac{i \pi}{2^{n+1}}}  , \sin^2 \pr{\frac{(i+1) \pi}{2^{n+1}}} }
,}
for $i=0\ldots2^{n}-1$, and holes $H_i=I_{n,i}$.
This is because the logistic map and the tent map are metrically conjugate through the nonlinear transformation $y=\sin^{2}\pr{\pi x /2}$.
The average escape rate is given by Eq.~\eqref{expected} with $N=2^n$ and $h_i=2^{-n}$ since the invariant density of the (closed) logistic map is $\rho(x)=\frac{1}{\pi\sqrt{x(1-x)}}$.

\section{Expanding maps with distortion}\label{sec.exception}


Here we investigate a generalization of the doubling map (sometimes referred to as the skewed doubling map) and observe that inequality \eqref{inequality} is reversed for sufficient skewness.
We consider the the one-dimensional map
\es{f(x)=\begin{cases} u x, &   0\leq x <1/u \\
v (x-1/u), &   1/u\leq x \leq 1, \end{cases}
\label{skew}}
where $v=u/(u-1)$, $u >1$, and $u\neq2$.
Map \eqref{skew} has a uniform invariant density $\rho(x)=1$ on the unit interval and shares the same binary symbolic dynamics as the doubling map for the partition $\{[0,1/u), [1/u,1]\}$, and hence hierarchy of periodic sequences.
However, the map does not have a constant piecewise expansion rate and its Lyapunov exponent is given by $\frac{1}{u}\ln u+ \frac{1}{v}\ln v$.

\begin{figure}[t]
\begin{center}
\includegraphics[scale=0.24]{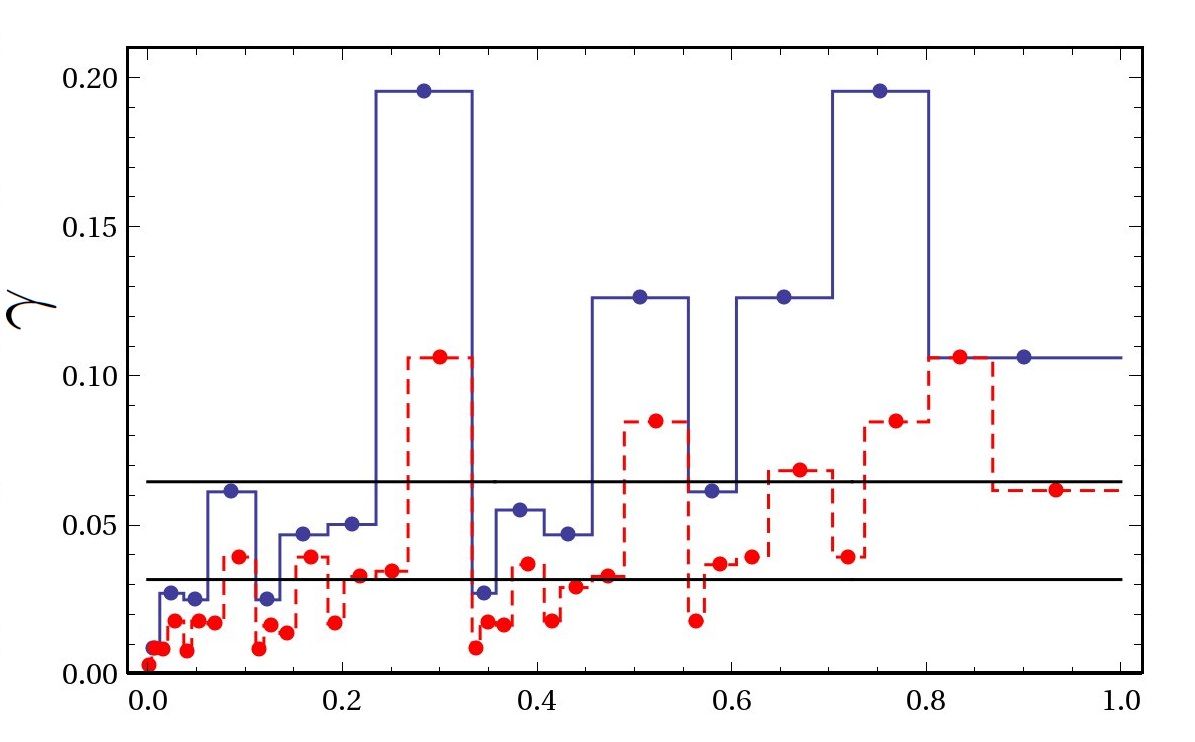}
\caption{\label{fig:skewedescape}  Escape rate as a function of hole position for Markov holes $H_i$ of different sizes $h_i$ for $n=4$ (blue) and $n=5$ (red, dashed) for the skewed map Eq.~\eqref{skew} with $u=3$.
The holes are centered at the dots and are of size equal to the corresponding tower widths.
The horizontal (black) lines correspond to $\bar{\gamma}_1$.}
\end{center}
\end{figure}

The natural Markov partition of $[0,1]$ under the dynamics of $f$ is composed of intervals of variable lengths $|I_{n,i}|\in \{ u^{-(n-l)}v^{-l} \big| l\in[0,n] \}$ with binomial occurrences.
Hence the corresponding Markov holes $H_i$ are not all of the same size and so $f$ is not a Fair Dice-Like (FDL) hyperbolic system \cite{B12}.
The escape rates for the case of $u = 3$ are plotted in Fig. \ref{fig:skewedescape} for $n=4,5$.
Notice that there is only one minimal escape rate at $H_0$.

The generalization of Eq.~\eqref{BY} for skewed maps (including skewed tent maps) is given by \cite{KL09}
\es{\tilde{\gamma}= h_i \pr{1- \Lambda_{x}^{-1} }+o(h_i)
,}
where $\Lambda_x=\frac{d}{dx} f^n(x)$ is the stability eigenvalue of the periodic orbit starting from $x\in H_i$, and is infinite if the orbit of $x$ is aperiodic.
Hence, the stability of a periodic orbit depends not only on its period but also on its symbolic sequence.
This implies that while the periodic orbit formula of  Eq.~\eqref{poform} can be naturally generalized to the current setting by replacing $2^{p}$ by $\Lambda_x$, the results of Sec \ref{ssec.sha} need to be reformulated in terms of stability orderings \cite{DM97} rather than just the period lengths.

\begin{figure}[t]
\begin{center}
\includegraphics[scale=0.3]{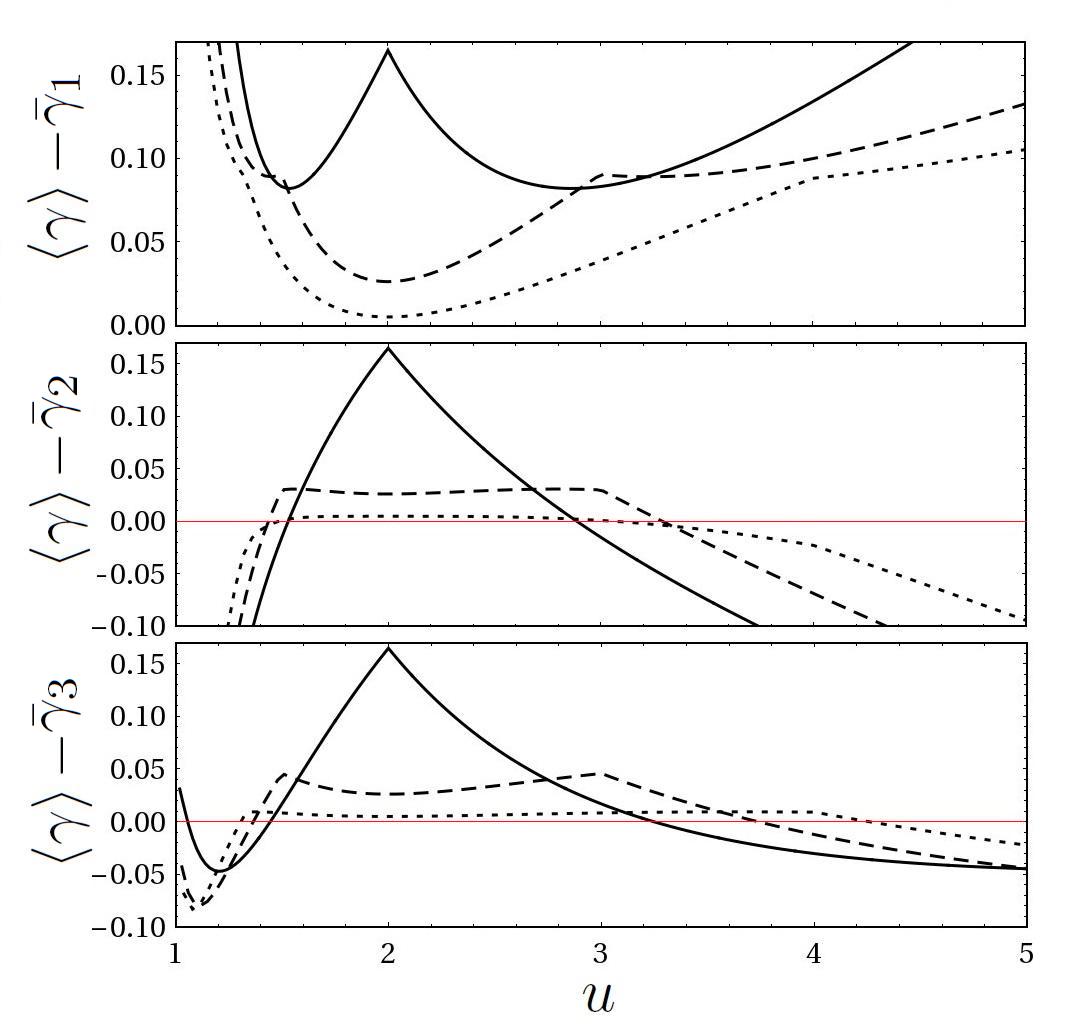}
\caption{\label{fig:gg} (Color online) A comparison between the average escape rate $\langle \gamma \rangle$ and the three naive estimates $(\bar{\gamma }_{1},\bar{\gamma }_{2},\bar{\gamma }_{3})$ as a function of $u$. The difference between the averages is plotted for $n=2,3,$ and $4$, corresponding to the full, dashed and dotted lines respectively. Notice that for $\bar{\gamma }_{2}$ and $\bar{\gamma }_{3}$, inequality \eqref{inequality} can be broken.
}
\end{center}
\end{figure}

For open skewed maps, one also needs to reconsider the ``naive'' estimate \eqref{naive}, as there now may exist several possible candidates.
We concentrate on the following three:
(i) Clearly, the average hole size is $2^{-n}$ (independently of $u$) and so a first candidate is simply $\bar{\gamma}_1=-\ln (1-2^{-n})$.
(ii) Assuming that the conditionally invariant density remains constant inside the open system, we obtain a second estimate which we denote as $\bar{\gamma }_{2}=  -\sum_{i=0}^{2^{n}-1} h_i \ln(1-h_i) = u^{-2n}(u-2u+2)^{n} + \ldots$ for $n\gg 1$, where we have used the fact that there are ${n \choose l}$ holes of size $u^{-(n-l)}v^{-l}$.
(iii) A third and not so naive estimate for the expected escape rate can be obtained from the random map setting, discussed in Sec.~\ref{sec.background}, by looking at the leading eigenvalue of $\bar{T}= \sum_{i=0}^{2^{n}-1} h_i T_i$.
For the case of $n=2$ this can be expressed in closed form as
\es{
\bar{\gamma}_{3}= -\ln \pr{\frac{ 3(u-1) + \sqrt{ (u-1) (4 u^{2}-7 u +7 )} }{2 u^2}}
.}
Note that estimates $\bar{\gamma}_{1}$, $\bar{\gamma}_{2}$ and $\bar{\gamma}_{3}$ are smooth functions of $u$ and for $u=2$ are equal to the original naive estimate \eqref{naive}.
Furthermore, we remark that the three estimates described above, contain increasingly more information about the system at hand with $\bar{\gamma}_{1}$ considering only the number of possible holes, $\bar{\gamma}_{2}$ also considering the variation in hole sizes, and finally $\bar{\gamma}_{3}$ also capturing some of the dynamics through the average transfer matrix $\bar{T}$.

A comparison between the exact value of $\langle \gamma \rangle$ obtained numerically from \eqref{expected} and the three candidate ``naive'' estimates is shown in Fig.~\ref{fig:gg} as a function of $u$ for $n=2,3,$ and $4$.
From the figure, it is clearly seen that the inequality is satisfied for case (i) entailing $\bar{\gamma}_{1} \leq \langle \gamma \rangle$ but not for cases (ii) and (iii) such that $\bar{\gamma}_{2,3}> \langle \gamma \rangle$ for sufficient distortion, i.e. when $u/v \gg 1$ or $u/v \ll 1$.
In contrast, the inequality in case (i) is strengthened with added distortion.
Therefore, the benchmark for comparison with $\langle\gamma \rangle$ may be of equal importance as the underlying dynamics of the map considered;
it might be interesting to investigate these further.
An alternative naive estimate would be to consider holes of constant size, i.e. non-Markov holes.

We briefly discuss the kinks observed in Fig.~\ref{fig:gg}.
For sufficiently large $u$, the eigenvalue associated with the fixed point at $1$ (equal to $1/v$), increases until it is equal to the eigenvalue associated with the whole repeller in the rest of the dynamics.
This degeneracy can only occur for holes which can break the transitivity of the dynamics and hence cause competing escape rates, in this case crossing over at the parameter value $u=n$.
By symmetry, the same occurs at $u=\frac{n}{n-1}$ (i.e $v=n$).
A similar effect was observed for the hyperbolic stadium billiard \cite{DG11} with two holes corresponding to a source and a sink which can be placed in such a way as to produce asymmetric transport.
In both cases, the presence of the hole renders the dynamics non-transitive.

\section{Discussion and conclusions}\label{sec.conclusions}

In this paper we have studied the dependence of the escape rate in uniformly hyperbolic dynamical systems on the location of Markov holes, with
focus on the ``typical'' escape rate.
We have analytically derived an exact periodic orbit formula which provides the escape rate as a function of the periodic orbits inside the hole.
Using an asymptotic expansion to second order in the hole size we obtained an expression for the average escape rate $\langle \gamma \rangle$ which is larger than the expectation~$\bar{\gamma}$ from the size of the hole $h$ alone, a surprising result in view of the fact that short periodic orbits typically lead to escape rates less than $\bar{\gamma}$.
We have shown however that the sheer number of holes with $\gamma_i > h$, in contrast to the much fewer but much more pronounced holes with $\gamma_i < h$, is what effectively causes the inequality $\langle\gamma\rangle \geq \bar{\gamma}$, with equality only in the limit $h\rightarrow0$.
Our results were illustrated and complemented with an exact numerical analysis and are valid for systems conjugate to the binary shift (e.g. linearly expanding maps, tent map, logistic map, and baker map).

An important question is the generality of these results to different classes of systems. In this regard, the skewed map discussed in
Sec.~\ref{sec.exception} shows that when the Markov partitions have different sizes, there are different possible generalizations of the naive estimate $\bar{\gamma}$ and that our main inequality $\langle \gamma \rangle \ge \bar{\gamma}$ holds only in one of the cases.
We have also performed numerical simulations in the chaotic diamond billiard, which is beyond the class of systems investigated here as there is no finite Markov partition of the phase space.
Our observations (not shown) indicate that if the hole is placed along the border of the billiard the inverse inequality i.e. $\langle \gamma \rangle \leq  \bar{\gamma}$ appears to hold for all hole sizes.
In general, it is an interesting open problem to verify in which classes of systems the inequality (or the reversed inequality)
holds systematically for all (small) holes sizes.

\begin{figure}[t]
\begin{center}
\includegraphics[scale=0.24]{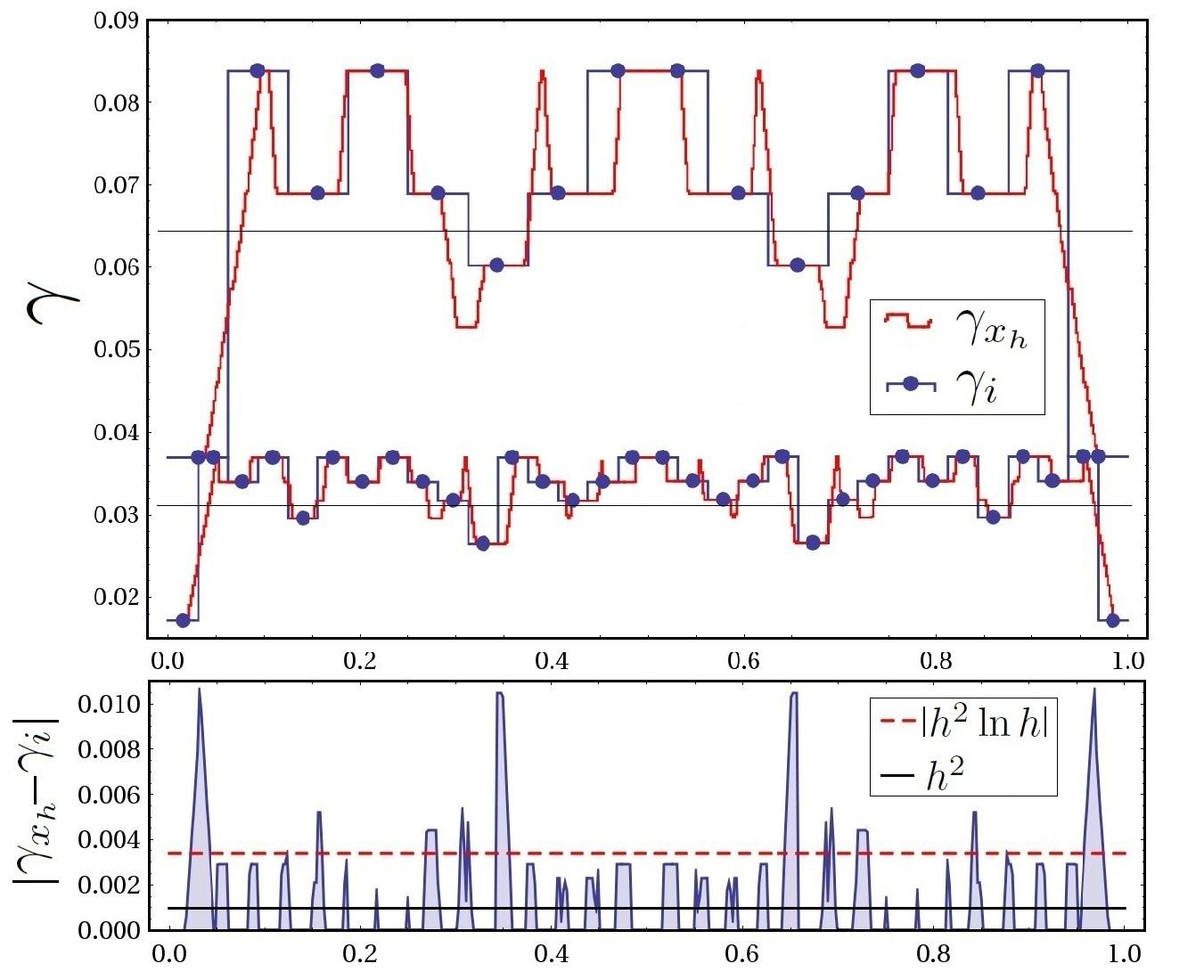}
\caption{\label{fig:nonmarkov} (Color online) \emph{Top:} The escape rate $\gamma_{x_h}$ for a hole of size $h=2^{-n}$ for $n=4$ (top curve) and $n=5$ (bottom curve) centered at $x_h\in[h/2,1-h/2]$ (red jagged curve) is compared with the escape rate $\gamma_i$ through $2^n$ Markov holes (blue line connecting dots).
The horizontal distance between two dots is equal to $h$.
\emph{Bottom:} Plot of $| \gamma_{x_h}-\gamma_i |$ for a hole of size $h=2^{-5}$. The discrete values of $\gamma_i$ seem to capture the main character of $\gamma_{x_h}$, modulo fluctuations of order $h^2 \ln h$.
}
\end{center}
\end{figure}

Another assumption in our analysis which asks for generalization is the choice of the holes to coincide with the Markov partitions of the
maps. It is thus natural to consider the doubling map with a non-Markov hole of size $h=2^{-n}$ centered at any point $x_h\in[h/2,1-h/2]$.
When $x_h$ is varied smoothly, the escape rate $\gamma_{x_h}$ becomes a highly non-smooth function which in the limit of $h\rightarrow0$
contains a dense set of maxima and minima; a fractal function~\cite{D11b}.
For small but finite holes however, $\gamma_{x_h}$ is expected to have many locally constant intervals.
This happens because for open chaotic systems, the largest invariant set that never reaches the hole is a
locally constant function of the hole size, shape and position~\cite{DW12}.
In Fig.~\ref{fig:nonmarkov} we have confirmed this picture for the doubling map by varying continuously the position of a hole of size $h = 2^n$ (we expect similar results also for hole sizes $h\neq2^n$).
Moreover, we observe that $\gamma_{x_h}$ is well described by the discrete $\gamma_i$'s corresponding to the $2^n$ Markov holes, modulo fluctuations which appear to be of order $h^{2} \ln h$.
We stress however that these fluctuations appear unrelated to the $h^2 \ln h$ terms predicted in the asymptotic expansions of Ref.~\cite{D11b} since Fig.~\ref{fig:nonmarkov} compares the exact escape rate $\gamma_{x_h}$ with the one through its nearest Markov hole.
Interestingly, we have observed numerically (not shown) that our basic inequality \eqref{inequality} remains valid if the average $\langle \gamma \rangle$ is computed using the continuum of possible hole positions (instead of only those at the Markov partitions).
Altogether, these results suggest that the Markov holes form a rigid ``Markov skeleton'' of the full curve and can be considered representative of the more elaborate problem.

Finally, it is worth comparing our results to different approaches. In Ref.~\cite{KGDK12} it was shown analytically that the average
{\it diffusion coefficient}, calculated over different hole positions in a related dynamical system, is exactly equal to the size of the
holes $h$.
In contrast, our results show that the escape rate is always larger than $h$.
Ref.~\cite{AB07} made the insightful suggestion to interpret the transfer matrix $T$ of our Markov maps as the adjacency matrix (also
called the connection matrix) of a network (or weighted directional graph).  In this analogy, making a hole in the phase space $\mathcal{M}$
is equivalent to consider one of the $N$ network vertices to be a ``sink'' (not able to transmit information)~\cite{AB07,AB10,BB11}.
Our results show that, for the strongly connected networks corresponding to Markov maps, the random choice of a sink leads to a loss of
information per unit time which decays typically faster than the expected $e^{-1/N}$.
It is also worth noting that, beyond the interpretation mentioned in the first paragraph of this paper, there are many different alternative
approaches to random leaks such as the randomly perturbed metastable interval maps considered in Ref. \cite{BV12}.

\section*{Acknowledgments}

The authors gratefully acknowledge helpful discussions with Leonid A. Bunimovich, Charo Del Genio, Rainer Klages, Georgie Knight, and Tam\'{a}s T\'{e}l.
CD would also like to thank the MPI-PKS for its kind hospitality during his visit to Dresden in May 2012.


\begin{thebibliography}{99}


\bibitem{Yorke79}
{G. Pianigiani, and J. A. Yorke}, Expanding maps on sets which are almost invariant: decay and chaos, Trans. American. Math. Soc. \textbf{252}, 351, (1979).

\bibitem{BB90}
{W. Bauer and G. F. Bertsch}, Decay of ordered and chaotic systems, Phys. Rev. Lett. \textbf{65}, 2213, (1990).

\bibitem{APT12}
{E. G. Altmann, J. S. E. Portela, and T. T\'{e}l}, Leaking chaotic systems, to appear in Rev. Mod. Phys. preprint  	arXiv:1208.0254, (2012).


\bibitem{PP97}
{V. Paar and N. Pavin}, Bursts in average lifetime of transients for chaotic logistic map with a hole, Phys. Rev. E, \textbf{55}, 4112, (1997).

\bibitem{AT09}
{E. G. Altmann, and T. T\'{e}l}, Poincar\'{e} recurrences and transient chaos in systems with leaks, Phys. Rev. E, \textbf{79}, 016204, (2009).

\bibitem{KL09}
{G. Keller, and C. Liverani}, Rare Events, Escape Rates and Quasistationarity: Some Exact Formulae, J. Stat. Phys, \textbf{135}, 3, (2009),

\bibitem{AB10}
{V. S. Afraimovich, and L. A. Bunimovich}, Which hole is leaking the most: a topological approach to study open systems, Nonlinearity, \textbf{23}, 643, (2010).

\bibitem{BY11}
{L. A. Bunimovich and A. Yurchenko}, Where to place a hole to achieve a maximal escape rate, Israel Journal of Mathematics, \textbf{182}, 229, (2011).

\bibitem{DG11}
{C. P. Dettmann and O. Georgiou}, Transmission and Reflection in the Stadium Billiard: Time-dependent asymmetric transport, Phys. Rev. E, \textbf{83}, 036212, (2011)

\bibitem{KGDK12}
{G. Knight, O. Georgiou, C. P. Dettmann and R. Klages}, Dependence of chaotic diffusion on the size and position of holes, Chaos, \textbf{22}, 023132, (2012).

\bibitem{DW12}
{M. F. Demers, and P. Wright}, Behavior of the Escape Rate Function in Hyperbolic Dynamical Systems, Nonlinearity, \textbf{25}, 2133, (2012).

\bibitem{FP12}
{A. Ferguson, and M Pollicott}, Escape rates for Gibbs measures, Ergodic Theory Dyn. Syst., \textbf{32}, 961, (2012).



\bibitem{DY06}
{M. F. Demers and L.-S. Young}, Escape rates and conditionally invariant measures, Nonlinearity, \textbf{19}, 377, (2006).




\bibitem{BG97}
{A. Boyarsky and P G\'{o}ra}, \textit{Laws of chaos: invariant measures and dynamical systems in one dimension}, Boston: Birkh\"{a}user, (1997).

\bibitem{Ulam64}
{S. M. Ulam}, \textit{Problems in Modern Mathematics}, (New York: Interscience), (1964).


\bibitem{GF07}
{G. Froyland}, On Ulam approximation of the isolated spectrum and eigenfunctions of hyperbolic maps, Discrete Contin. Dyn. Syst, \textbf{17}, 671, (2007).

\bibitem{WB11}
{W. Bahsoun and C. Bose}, Invariant Densities and Escape Rates: Rigorous and Computable Approximations in The $L^{\infty}$-norm, Nonlinear Analysis, \textbf{74}, 4481, (2011).


\bibitem{BFGTM12}
{C Bose, G. Froyland, C. G.Tokman and R. Murray}, Ulam's method for Lasota-Yorke maps with holes, preprint arXiv:1204.2329, (2012).

\bibitem{BD07}
{L. A. Bunimovich and C. P. Dettmann}, Peeping at chaos: Nondestructive monitoring of chaotic systems by measuring long-time escape rates, Europhys. Lett, \textbf{80}, 40001, (2007).

\bibitem{D11}
{C. P. Dettmann}, Recent advances in open billiards with some open problems, in \textit{Frontiers in the study of chaotic dynamical systems with open problems}, (Ed. Z. Elhadj and J.C. Sprott), World Sci. Publ., (2011).


\bibitem{AE11}
{E. G. Altmann and A. Endler}, Noise-Enhanced Trapping in Chaotic Scattering, Phys. Rev. Lett. \textbf{105}, 244102, (2010).


\bibitem{SP84}
{S. Pelikan}. Invariant densities for random maps for the interval, Trans. Amer, \textbf{281}, 813, (1984).

\bibitem{LPPV96}
{V. Loreto, G. Paladin, M. Pasquini and A. Vulpiani}, Characterization of chaos in random maps, Physica A, \textbf{232}, 189, (1996).






\bibitem{Solovev66}
{A. D. Solov'ev}, c.  Theor. Prob. Appl. {\bf 11}, 276 (1966).

\bibitem{BT82}
{G. Blom and D. Thorburn}, How many random digits are required until given sequences are obtained? J. Appl. Prob. {\bf 19}, 518 (1982).

\bibitem{AAC90}
{R. Artuso, E. Aurell and P. Cvitanovi\'c}, Recycling of strange sets: I. Cycle expansions, Nonlinearity {\bf 3}, 325, (1990).


\bibitem{chaos}
{P. Cvitanovi\'c et al}, \textit{Chaos: Classical and Quantum}, Copenhagen: Niels Bohr Institute, \textsc{ChaosBook.org}  (2009).


\bibitem{D11b}
{C. P. Dettmann}, Open circle maps: Small hole asymptotics.  arxiv:1112.5390 (2011).

\bibitem{OEIS}
{OEIS} The online encyclopedia of integer sequences, {\tt http://www.oeis.org} (2010).

\bibitem{MPS10}
{R. Markarian, E. J. Pujals, and M. Sambarino}, Pinball billiards with dominated splitting, Ergod. Th. \& Dynam. Sys., \textbf{30}, 1757, (2010).



\bibitem{BV87}
{N. L. Balazs and A. Voros}, The Quantized Baker's transformation, Europhys. Lett., \textbf{4}, 1089, (1987).

\bibitem{NM05}
{S. Nonnenmacher and M. Zworski}, Fractal Weyl laws in discrete models of chaotic scattering, J. Phys. A :Math. Gen, \textbf{38}, 10683, (2005).



\bibitem{B12}
{L. A. Bunimovich}, Fair dice-like hyperbolic systems. Dynamical Systems and Group Actions, \textbf{567}, 79. (2012).


\bibitem{DM97}
{C. P. Dettmann, and G. P. Morriss}, Stability ordering of cycle expansions, Phys. Rev. Lett. \textbf{78}, 4201, (1997).



\bibitem{AB07}
{V. S. Afraimovich and L. A. Bunimovich}, Dynamical networks: interplay of topology, interactions and local dynamics, Nonlinearity, \textbf{20}, 1761, (2007).

\bibitem{BB11}
{Y. Bakhtin, and L. A. Bunimovich}, The optimal sink and the best source in a Markov chain, J. Stat. Phys., \textbf{143}, 943, (2011).



\bibitem{BV12}
{W. Bahsoun, and S. Vaienti}, Escape Rates Formulae and Metastablilty for Randomly perturbed maps, preprint 	arXiv:1206.3654v2 , (2012).




\end{thebibliography}
\end{document}